\documentclass[aps,prl,twocolumn,superscriptaddress,groupedaddress,10pt]{revtex4}

\usepackage{amsmath}
\usepackage{amssymb}
\usepackage{graphicx}
\usepackage{epsfig}
\usepackage{dcolumn}
\usepackage{bm}
\usepackage{bm}
\usepackage{blindtext}
\usepackage{natbib}
\usepackage{booktabs}
\usepackage{mathrsfs}
\usepackage{epstopdf}
\usepackage{color}
\def\be{\begin{equation}}
\def\ee{\end{equation}}
\def\bea{\begin{eqnarray}}
\def\eea{\end{eqnarray}}

\begin{document}

\title{Quantum criticality of the Ising-like screw chain antiferromagnet SrCo$_2$V$_2$O$_8$ in
a transverse magnetic field}

\author{Y. Cui}
\affiliation{Department of Physics and Beijing Key Laboratory of
Opto-electronic Functional Materials $\&$ Micro-nano Devices, Renmin
University of China, Beijing, 100872, China}

\author{H. Zou}
\affiliation{Tsung-Dao Lee Institute $\&$ School of Physics and Astronomy,
Shanghai Jiao Tong University, Shanghai, 200240, China}

\author{N. Xi}
\affiliation{Department of Physics and Beijing Key Laboratory of
Opto-electronic Functional Materials $\&$ Micro-nano Devices, Renmin
University of China, Beijing, 100872, China}

\author{Zhangzhen He}
\email{hezz@fjirsm.ac.cn}
\affiliation{State Key Laboratory of Structural Chemistry, Fujian Institute
of Research on the Structure of Matter, Chinese Academy of Sciences, Fuzhou,
Fujian 350002, China}

\author{Y. X. Yang}
\affiliation{State Key Laboratory of Surface Physics, Department of Physics, Fudan University, Shanghai 200433, China}

\author{L. Shu}
\affiliation{State Key Laboratory of Surface Physics, Department of Physics, Fudan University, Shanghai 200433, China}
\affiliation{Collaborative Innovation Center of Advanced Microstructures, Nanjing 210093, China}

\author{G. H. Zhang}
\affiliation{Department of Physics and Beijing Key Laboratory of
Opto-electronic Functional Materials $\&$ Micro-nano Devices, Renmin
University of China, Beijing, 100872, China}

\author{Z. Hu}
\affiliation{Department of Physics and Beijing Key Laboratory of
Opto-electronic Functional Materials $\&$ Micro-nano Devices, Renmin
University of China, Beijing, 100872, China}

\author{T. Chen}
\affiliation{Department of Physics and Beijing Key Laboratory of
Opto-electronic Functional Materials $\&$ Micro-nano Devices, Renmin
University of China, Beijing, 100872, China}

\author{Rong Yu}
%\email{rong.yu@ruc.edu.cn}
\affiliation{Department of Physics and Beijing Key Laboratory of
Opto-electronic Functional Materials $\&$ Micro-nano Devices, Renmin
University of China, Beijing, 100872, China}

\author{Jianda Wu}
\email{wujd@sjtu.edu.cn}
\affiliation{Tsung-Dao Lee Institute $\&$ School of Physics and Astronomy,
Shanghai Jiao Tong University, Shanghai, 200240, China}

\author{Weiqiang Yu}
\email{wqyu_phy@ruc.edu.cn}
\affiliation{Department of Physics and Beijing Key Laboratory of
Opto-electronic Functional Materials $\&$ Micro-nano Devices, Renmin
University of China, Beijing, 100872, China}

%\date{\today}
%\pacs{74.70.-b, 76.60.-k}

\begin{abstract}

The quantum criticality of an Ising-like screw chain antiferromagnet SrCo$_2$V$_2$O$_8$,
with a transverse magnetic field applied along the crystalline $a$-axis,
is investigated by ultra-low temperature NMR measurements.
The N\'{e}el temperature is rapidly and continuously suppressed by the field,
giving rise to a quantum critical point (QCP) at $H_{C{_1}}$$\approx$~7.03~T.
Surprisingly, a second QCP at $H_{C{_2}}\approx$~7.7~T featured with gapless excitations is resolved from
both the double-peak structure of the field dependent spin-lattice relaxation rate $1/^{51}T_1$ at low temperatures
and the weakly temperature-dependent $1/^{51}T_1$ at this field.  Our data,
combined with numerical calculations, suggest that the induced effective staggered transverse field
significantly lowers the critical fields, and leads to an exposed QCP at $H_{C{_2}}$, which belongs to
the one-dimensional transverse-field Ising universality.

\end{abstract}

\maketitle

Novel quantum states with exotic excitations can emerge near a quantum phase
transition~\cite{Sachdev__2011,Loehneysen_JLTP__2010,Coleman_nature__2005}.
As one prototype for quantum phase transitions and quantum criticality, the
one-dimensional transverse-field Ising model (1DTFIM)
has been widely investigated ~\cite{Pfeuty_AP__1970,Barouch_PRA_1971, Doniach_PRB_1978,Nijhoff_PA_1984,
Zamolodchikov_IJMPA__1989,Tsvelik_PRL_2003,Perk_JSP__2009,WuJD_PRL_2014,WuJD_PRB_2018}.
In this model, the magnetic order is suppressed by a transverse magnetic field,
resulting in an order-disorder quantum phase transition,
with gap closed at a quantum critical point (QCP), $H_C$.
It has been predicted that a quantum $E_8$ integrable model emerges
near the vicinity of the QCP when a weak longitudinal field is applied
~\cite{Zamolodchikov_IJMPA__1989}.
The experimental realization of such a QCP is yet challenging because
the interchain exchange couplings in real materials usually
stabilize the magnetic order at $T>$ 0 and mask
the genuine 1D quantum criticality. Nonetheless, a 1DTFIM-like QCP
hidden in the 3D ordered phase was suggested in the ferromagnetic
chain compound CoNb$_2$O$_6$~\cite{Imai_PRX_2014,Coldea_Science_2010,Balents_NP_2010}.

\begin{figure}[t]
\includegraphics[width=7cm,height=6cm]{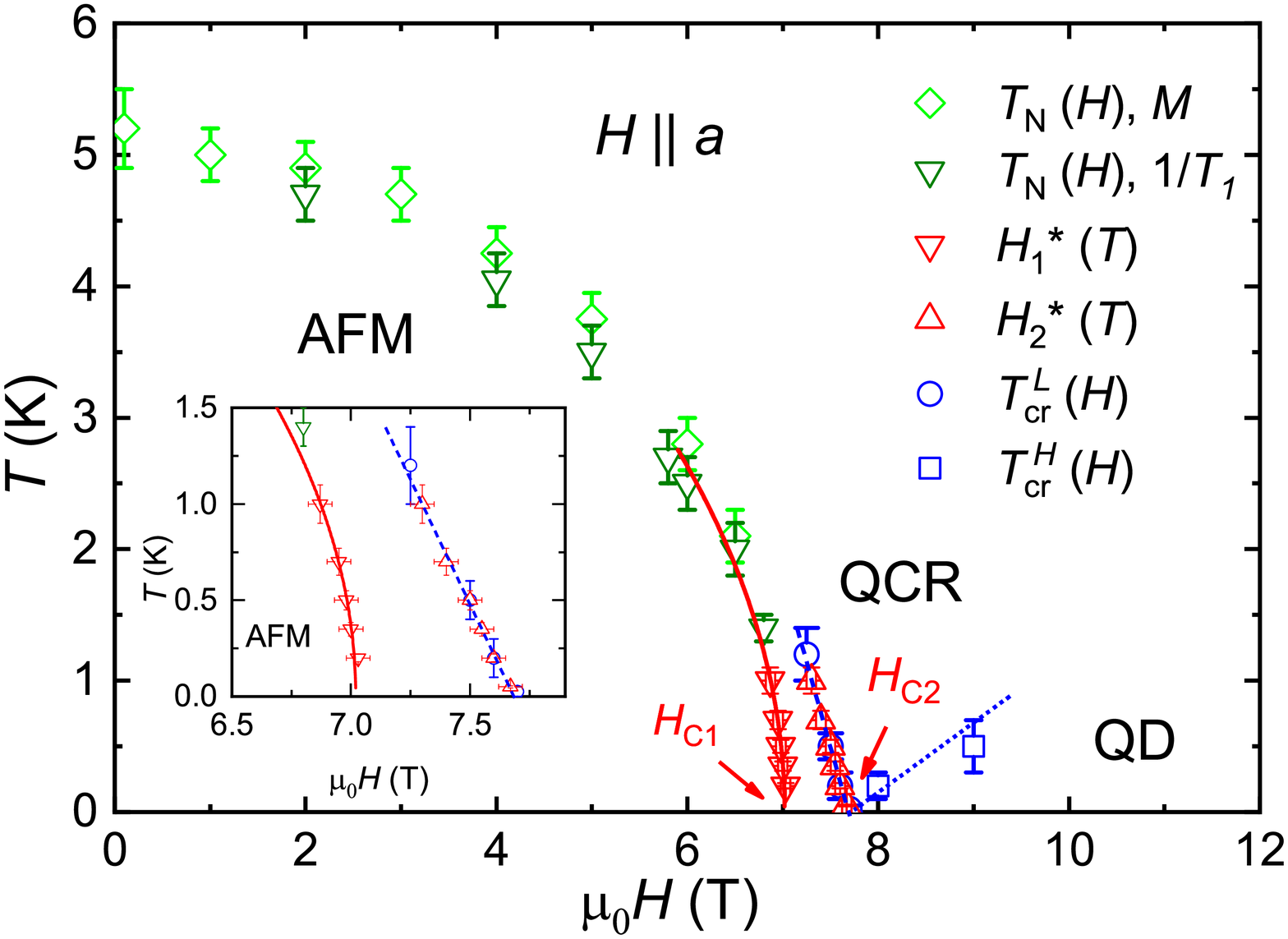}
\caption{\label{phasediagram} Phase diagram of SrCo$_2$V$_2$O$_8$ in a transverse
field along the $a$-axis. AFM, QCR and QD refer to the antiferromagnetic phase,
the quantum critical regime, and the quantum disordered regime.
Diamonds represent the N\'{e}el temperatures $T_N$ determined by the
magnetization data.  Down triangles, circles, and squares represent
$T_N$, the crossover temperatures
$T_{cr}^L$ and $T_{cr}^H$, respectively, all determined by the $1/^{51}T_1$
with different schemes of measurements (main text).
The solid line is a function fit to $T_N\sim$$(H_{C1}-H)^{0.5}$ with $H_{C1}\approx$ 7.03~T.
The blue lines denote function fits to $T_{cr}^{L,H}\sim$$|H_{C2}-H|$ with
$H_{C2}\approx$~7.7~T. Inset: An enlarged view of the fitting near the two QCPs.}
\end{figure}

\begin{figure}[t]
\includegraphics[width=8.5cm,height=6.5cm]{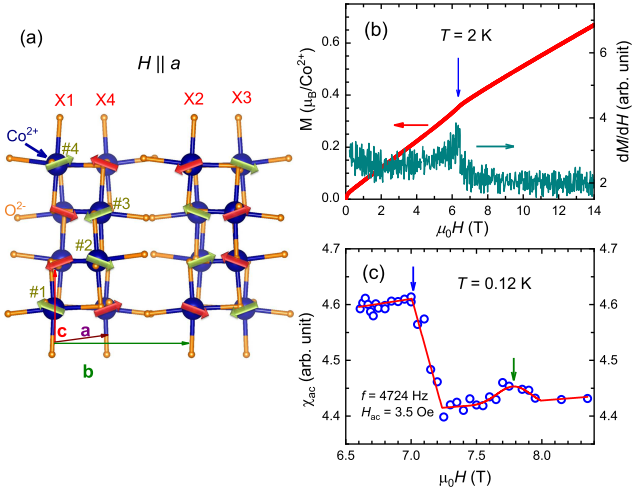}
\caption{\label{struc}(a) A schematic drawing of a unit cell of SrCo$_2$V$_2$O$_8$.
$a$, $b$ and $c$ label the tetragonal axes, X1 to X4 label
four neighboring spin chains.
$\#1$ to $\#4$ label the Co$^{2+}$ sites along one screw chain.
The arrows placed on the Co$^{2+}$ sites
mark the direction of the effective staggered field created
by a $dc$ field applied along the $a$-axis.
(b) The $dc$ magnetization $M$ vs. $H$ at 2~K and
its derivative with the field. The arrow marks
a peaked feature in $dM/dH$.
(c) The  $ac$ susceptibility $\chi_{ac}$ as a function of
$H$ at 0.12~K. The arrows mark a sharp drop and a
round peak in the data, respectively.}
\end{figure}

Recently, a class of Ising anisotropic antiferromagnetic
screw chain compounds $A$Co$_2$V$_2$O$_8$($A=$Ba,Sr)~\cite{He_PRB_2005,He_PRB_2006}
attracts a lot of research attention.
In these compounds, the easy-axis is along the chain direction
($c$-axis), and the intrachain exchange coupling
$J_z\approx$ 5-7~meV~\cite{Itoh_PRL_2007,Bera_PRB__2017}.
Both compounds order at fairly low temperatures (about 5~K),
which implies the interchain exchange coupling $J^\prime\ll J$~\cite{Itoh_PRL_2008,Bera_PRB__2014,Klanjsek_PRB__2015}.
The magnetic order can be suppressed by a transverse field applied
in the $ab$-plane~\cite{He_PRB_2005,He_PRB_2006}.
Interestingly, the critical field values and the quantum critical behaviors are
very different for fields applied along the [100]
and the [110] direction~\cite{Kimura_JPSJ__2013,Lorenz_PRB__2013,
Okutani_2015,Wang_PRB__2016, Bera_PRB__2017,Matsuda_PRB_2017, Faure_np__2018,Wang_PRL__2018}.
In BaCo$_2$V$_2$O$_8$,
separated 3D and 1D QCPs were suggested with field along the [110] direction at
$H\gtrsim$~20~T ~\cite{Wang_PRL__2018}.
For the field along the [100] direction, $H_{C}\sim$~10~T~\cite{Lorenz_PRB__2013},
much lower than the $H_C$ along the [110] direction.
With field along the [100] direction, the linear confining potential
arising from the interchain correlation at fields below $H_C$ and the paramagnetism
at fields above $H_C$ create different topological excitations
~\cite{Faure_np__2018,Giamarchi_arxiv_180706029}.
The critical behavior near $H_C$ is, however, not investigated.

In order to reveal the properties in the critical regime,
low-temperature and low-energy probes are desired.
Here we report our susceptibility and $^{51}$V NMR measurements on
SrCo$_2$V$_2$O$_8$ with $H$$\parallel$~$a$ and temperatures down to 50~mK.
The main results are summarized in the phase diagram
of Fig.~\ref{phasediagram}. From the double-peak signature of the
$1/^{51}T_1$ data, we identify two QCPs at $H_{C1}\approx$~7~T
and $H_{C2}\approx$~7.7~T, respectively.
By analyzing the NMR line splitting at low temperatures, we show
that the system establishes three dimensional (3D) magnetic ordering below $H_{C1}$
and the paramagnetic phase above $H_{C2}$.
In between $H_{C1}$ and $H_{C2}$, the line splitting is
absent. We observe $1/^{51}T_1\sim T^{2.2}$ at $H_{C1}$ and
becomes almost temperature-independent at $H_{C2}$, signaling
gapless excitations at these two QCPs.
Our experiment reveals highly nontrivial critical behaviors of the system,
and hence discovers a new route in accessing novel quantum
criticality in quasi-1D Ising-like antiferromagnets, accomplished from screwed lattice structures.

\begin{figure}[t]
\includegraphics[width=8.5cm,height=6.5cm]{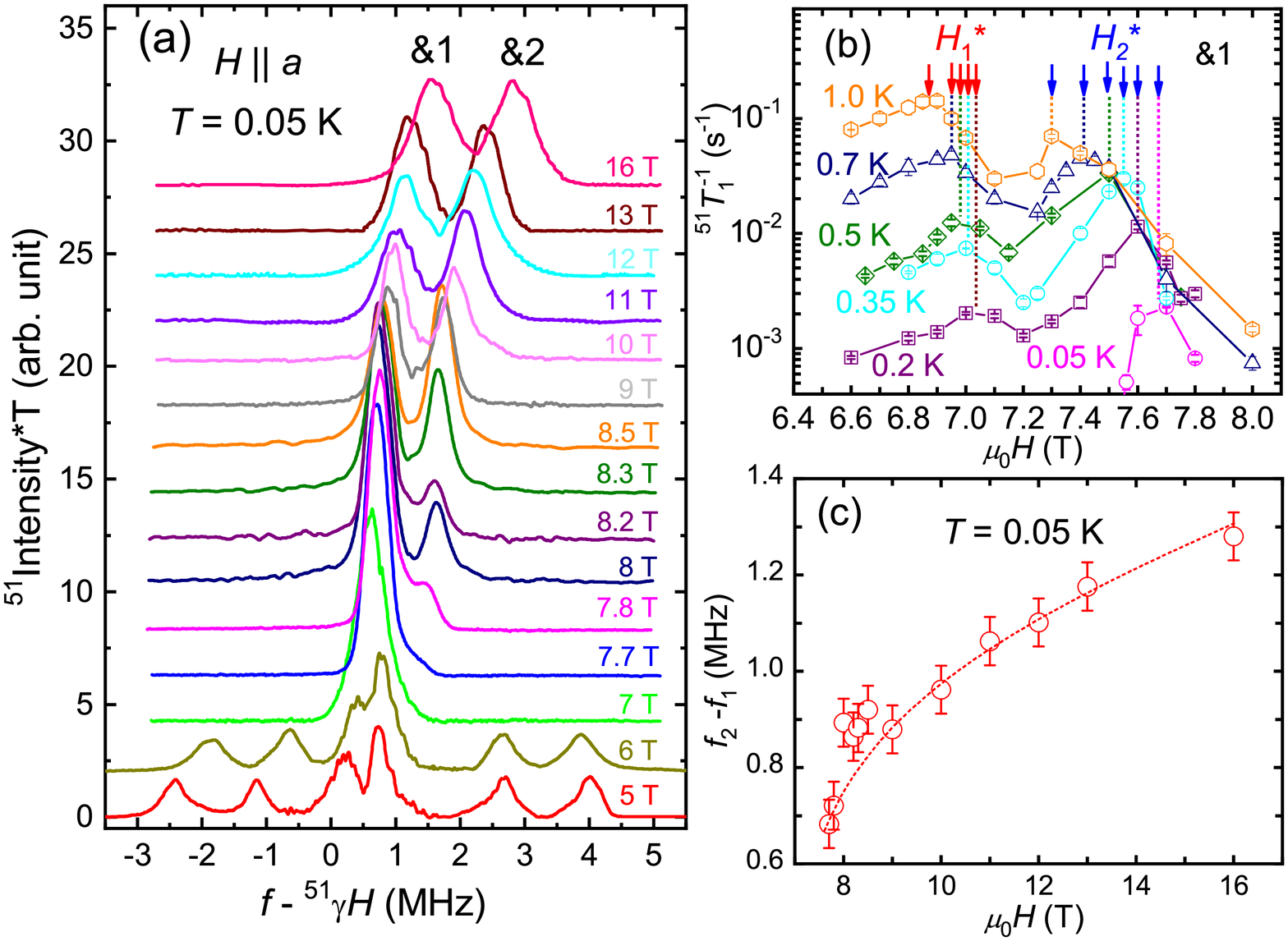}
\caption{\label{spec} (a)The $^{51}$V NMR spectra measured at 0.05~K with typical
fields. Escalated vertical offsets are applied for clarity.
$\&1$ and $\&2$ mark the two resonance peaks at fields above 7.7~T.
(b) The field-dependence of $1/^{51}T_1$ at fixed temperatures
measured on peak $\&1$. $H_1^*$ and $H_2^*$ mark the field locations
of peaked $1/^{51}T_1$ at each temperature. (c) The frequency difference
between peak $\&1$ and $\&2$ as a function of field.}
\end{figure}

\begin{figure*}[t]
\includegraphics[width=16cm,height=6cm]{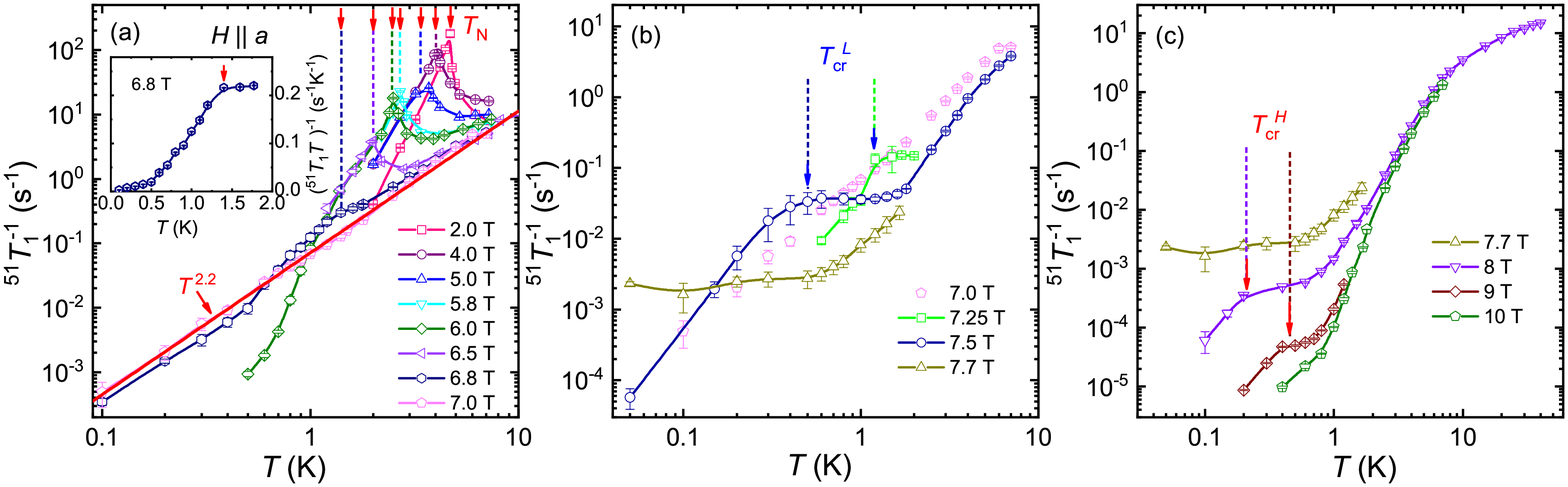}
\caption{\label{slrr}The spin-lattice relaxation rate $1/^{51}T_1$ as functions of temperatures
at various fields. (a) Data at fields below 7~T.  The red arrows mark the peaked
feature at $T_N$. The red straight line is a power-law function fit to the data at 7~T
with $1/T_1$$\sim$$T^\alpha$ ($\alpha$=2.2).
Inset: $T_N$ at 6.8~T resolved by a sharp drop of $1/^{51}T_1T$ upon cooling.
(b) Data at fields from 7~T to 7.7~T. (c) Data at fields above 7.7~T.
$T_{cr}^L$ ($T_{cr}^H$) at each field marks a
crossover temperature between a gapped and a gapless behavior.}
\end{figure*}

The $dc$ magnetization was measured above 2~K and the $ac$ susceptibility was measured at 0.12 K.
Details on sample preparation and NMR measuring techniques are described in Sec.~S1~\cite{SM}.
The $^{51}$V ($I=7/2$, $^{51}\gamma$ =11.198 MHz/T) NMR was performed
in a dilution refrigerator with the spin-echo technique.
The spin-lattice relaxation rate $1/^{51}T_1$ was measured by the inversion-recovery method,
where the nuclear-magnetization is well fit with the recovery function
for $I=7/2$ spins~\cite{MacLaughlin_PRB__1971_4,Fujita1984} without any
stretching behavior, indicating high-quality of the sample (Sec.~S8~\cite{SM}).

SrCo$_2$V$_2$O$_8$ crystalizes in a tetragonal space group $I4_1cd$~\cite{He_PRB_2006}
as shown in Fig.~\ref{struc}(a). It consists of screwed chains along the c-axis with alternative clockwise
and anticlockwise screwing directions among neighboring chains~\cite{Bera_PRB__2014}.
Each unit cell contains four chain segments in the $ab$-plane (labeled as X1 to X4),
and each of them further includes four sites labeled as $\#1$ to $\#4$.
Similar to BaCo$_2$V$_2$O$_8$, the primary interactions can be formulated as
weakly coupled effective spin-$1/2$ antiferromagnetic XXZ chains with strong Ising anisotropy
(Sec.~S2~\cite{SM}). Because the vertical O-Co-O bond
of SrCo$_2$V$_2$O$_8$ in the local CoO$_6$ octahedral tilts from the $c$-axis
by $\sim 5.1^\circ$ with a four-site period~\cite{Lorenz_PRB__2014},
the applied $a$-axis transverse field $H$
induces a two-period staggered field $H_y$ and a
four-period field $H_z$ along the chain~\cite{Kimura_JPSJ__2013},
as demonstrated in Fig.~\ref{struc}(a).

The applied and staggered fields suppress the antiferromagnetic Ising order~\cite{Kimura_JPSJ__2013},
leading to a phase transition, signaled in our magnetization data.
The N\'{e}el temperatures determined from the magnetization
data at different fields (Sec. S3~\cite{SM}) are plotted
in Fig.~\ref{phasediagram}.
At 2~K, the $dc$ magnetization is shown as a function of field in Fig.~\ref{struc}(b),
which exhibits a kinked feature at about 6.4~T. The $dM/dH$ clearly
resolves a peak at 6.4~T, consistent with a field-induced phase transition as reported
earlier~\cite{He_PRB_2006,Bera_PRB__2014}. The $ac$ susceptibility data at 0.12~K,
as shown in Fig.~\ref{struc}(c), demonstrate a sharp drop at $\mu_0H\approx$~7~T,
indicating a quantum phase transition. In addition, we also observe
a weak round peak at $\mu_0H\approx$ 7.7~T, indicating a crossover rather than a phase transition, which is a
precursor to a 1D QCP and will be discussed in more details later.
In the following, we show unambiguous evidence
for distinctive magnetic behaviors separated by 7~T and 7.7~T by the $^{51}$V
NMR spectra and the spin-lattice relaxation rate $1/^{51}T_1$.

The ground state at each field can be identified by the NMR spectra at
low temperatures. The $^{51}$V spectra at frequencies
close to $^{51}\gamma H$ are measured at typical fields and
shown in Fig.~\ref{spec}(a). Three regimes are clearly seen.
Below 7~T, the $^{51}$V spectra
have six peaks. Such a line splitting
is caused by twinned magnetic domains in the Ising ordered phase
originating from frustrated interchain exchange couplings~\cite{Kawasaki_JPSJ__2014, SM,Bao_CPL_2016,Shen_NJP_2019};
In the intermediate field range from 7~to 7.7~T, a single NMR
line is seen close to the central frequency $f\approx$$^{51}\gamma$$H$,
which suggests the absence of magnetic ordering;
At fields above 7.7~T, two NMR lines, labeled as $\&1$ and $\&2$,
emerge and are resolvable at temperatures above 10~K (Fig. S8~\cite{SM}).
The frequency difference of the two peaks increases with
field (Fig.~\ref{spec}(c)), implying the system is
paramagnetic when the field is above 7.7 T (Sec. S7~\cite{SM}).

The  magnetic transition at each field can be resolved
by $1/^{51}T_1$, which is measured at the primary peak of the spectra.
In Fig.~\ref{slrr}(a)-(c), the $1/^{51}T_1$ with increasing temperature is shown
at various fields. For fields below 6.5~T, a peak is observed in $1/^{51}T_1$, which
determines the magnetic ordering temperature $T_N$. At 6.8~T, although
a sharp drop of $1/T_1$ is not clearly discernible, the magnetic
transition is resolved by a sudden drop of $1/^{51}T_1T$
at 1.35~K, as demonstrated in the inset of Fig.~\ref{slrr}(a).
For any fields above 7~T, no peaked feature in the $1/^{51}T_1$
is observed, indicating
the absence of magnetic ordering.
The determined $T_N$ with the fields up to 6.8~T is
shown in Fig.~\ref{phasediagram}.
It agrees well with $T_N$ determined from our
$dc$ susceptibility data (Sec.~S3~\cite{SM}).

To determine the transition precisely, we perform a field scan of
$1/^{51}T_1$ at fixed $T$ with increasing field, and the result is
shown in Fig.~\ref{spec}(b).
Surprisingly, at low temperatures the $1/^{51}T_1$ curve shows two prominent
peaks with field. For instance, at 0.2~K, a low-field peak is located at $\sim$7~T
(marked as $H_1^*$), while a high-field one is at $\sim$7.6~T
(marked as $H_2^*$). These are close to the critical and the crossover field
revealed in the $ac$ susceptibility, respectively.
We then plot $H_1^*(T)$ and $H_2^*(T)$ as functions of temperatures in the phase diagram
of Fig.~\ref{phasediagram}.

The $H_1^*(T)$ line, together with the high-temperature
$T_N$ data, portray the critical trajectory, which is well fit
by $T_N\sim (H-H_{C1})^\phi$ with $H_{C1}$= 7.03~T and a critical exponent
$\phi\approx$~0.5$\pm$0.09.
Taking into account that the absence of the NMR line splitting,
the peaked feature in the $1/^{51}T_1(H)$, the sharp drop of $\chi_{ac}$,
and the mean-field-like exponent $\phi\approx0.5$, all emerge at the same
field $H_{C1}\approx$ 7~T, we
conclude that $H_{C1}$ is a (3+1)D Ising critical point.

The determined $T$-$H$ curve from $H_2^*(T)$ follows a straight line (Fig.~\ref{phasediagram}).
Given that no singularity of any thermodynamic quantity was observed,
we understand it as a crossover approaching to a 1D QCP at $H_{C2}\approx$ 7.7~T.
This $H_{C2}$ is also consistently resolved by the temperature evolution of $1/^{51}T_1$
data at different fields (Fig.~\ref{slrr}).
For fields either below or above $H_{C2}$, upon warming, the $1/^{51}T_1$
increases rapidly, and then crossovers to a weakly temperature-dependent behavior.
This clearly indicates a gapped behavior below the crossover temperature
$T_{cr}^L$ (for $H<H_{C2}$) or $T_{cr}^H$ (for $H>H_{C2}$) (Fig.~\ref{slrr}(b)-(c)).
$T_{cr}^L$ decreases with field, while $T_{cr}^H$ increases with field.
Interestingly, $T_{cr}^L$ overlaps with the crossover line $H_2^*(T)$ in the
studied temperature regime (Fig.~\ref{phasediagram}).
It scales linearly with field by
$T_{cr}^L\sim |H-H_{C2}|$, implying the critical exponents $\nu z=1$, consistent
with the 1DTFIM universality class~\cite{Sachdev__2011}.

In the following, we show how the transitions at these two QCPs are affected by the
induced staggered fields. First, the existence of a 1D QCP at a low field is
further supported by our theoretical calculation. We model the primary interactions of
SrCo$_2$V$_2$O$_8$ to a spin-$1/2$ Heisenberg-Ising chain, and
the effective Hamiltonian reads as,
\bea
\hspace{-9mm}&&H = \sum\limits_i {J[S_i^xS_{i + 1}^x + S_i^yS_{i + 1}^y + \Delta S_i^zS_{i + 1}^z]} -  \nonumber \\
\hspace{-9mm} &&{\mu _B}{g_x}\sum\limits_i {[HS_i^x + {H_y}{{( - 1)}^i}S_i^y + {H_z}S_i^z\cos (\pi (2 i - 1)/4)]}
 \label{hamiltonian}
\eea
with $J\approx$~3.7 meV and the anisotropic factor $\Delta\approx$~2.1~\cite{Bera_PRB__2014, Kimura_JPSJ__2013, Wang_PRB__2015,Wang_PRB__2016}.
Here $g_x$ is the gyromagnetic ratio,  $H$ is the applied field along the $a$-axis, $H_y$ and $H_z$ are, respectively,
induced two- and four-period staggered fields along the $b$- and $c$-axes,
due to screwed lattice structures~\cite{Kimura_JPSJ__2013}.
By fitting the high-field magnetization data~\cite{Okutani_PP_2015},
we deduced $g_x\approx$~3.7, $H_y\approx$~0.29~$H$, and $H_z\approx$~0.14~$H$.
We then apply the infinite time evolving block decimation (iTEBD) method~\cite{Vidal_PRB_2018}
to determine the local static moment with respect to the field at zero temperature,
shown in Sec.~S4~\cite{SM}. The long-range antiferromagnetic order is continuously suppressed
to zero at $H_C\approx$~7.6~T. Note that the calculated $H_C$ value is very close
to the measured $H_{C2}$.
Compared to $H\parallel$~[110], the $H_C$ for $H\parallel$~[100]
is largely reduced~\cite{Lorenz_PRB__2014}. This is primarily caused by the induced staggered
field $H_y$, which is the strongest for $H\parallel$~[100] and tuned down to
zero for $H\parallel$~[110]~\cite{Lorenz_PRB__2014}, consistent
with the theoretical model~\cite{Kimura_JPSJ__2013}.

Secondly, the scaling of the order parameter at $H_C$ from the iTEBD calculation finds
the critical exponent $\beta\approx$1/8 (Fig.~S6~\cite{SM}).
This indicates that though $H_C$ can be substantially reduced by
the staggered field $H_y$, the QCP is still governed by the
1DTFIM universality~\cite{Giamarchi_arxiv_180706029,SM} as described above.

Lastly, in quasi-1D systems, the locations of 1D and 3D QCPs
reflect the competition between the effective interchain
couplings and the external and induced fields,
and are usually strongly material dependent.
For example, the 1D QCP of CoNb$_2$O$_6$ is hidden inside a magnetically ordered phase
close to a 3D QCP~\cite{Imai_PRX_2014}. In BaCo$_2$V$_2$O$_8$ with a
transverse field applied along the [110] direction, a 3D QCP
is located at about 20~T, succeeded by a 1D QCP at about 40~T~\cite{Wang_PRL__2018}.
For SrCo$_2$V$_2$O$_8$, given its similar structural and magnetic properties to
BaCo$_2$V$_2$O$_8$, two QCPs with field along the [110]
direction are expected although not explored.
By rotating the field from the [110] to the [100] direction,
the induced staggered field might significantly lower both QCPs,
and our measurements unambiguously show that the 1D QCP
is outside the 3D ordered phase, yet are separated by a
small field of 0.7~T.

The observation of the 1D QCP outside the 3D ordered phase
prompts that the effective interchain couplings are very weak.
This is likely caused by the heavy frustration among the interchain
couplings~\cite{Bera_PRB__2014,Wang_PRB__2016}. The induced staggered
fields, which are tied up with the screwed structure, may further enhance the frustration effects and
help weaken the 3D order. However, a full discussion on this would need
detailed information on the interchain couplings,
which is beyond the scope of this paper.

In what follows, we discuss the spin dynamics at both the 1D and 3D
QCPs. As for the 3D QCP at $H_{C1}$, our $1/^{51}T_1$ data
exhibit a power-law scaling $1/^{51}T_1\sim T^{2.2}$ from 10~K down
to 0.1~K (Fig.~\ref{slrr}(a)). Such a power-law
scaling over two decades of temperatures
evidences gapless excitations at $H_{C1}$.
At the 1D QCP, $H_{C2}$, the $1/^{51}T_1$ stays nearly constant,
for over one decade in temperature below 0.5~K, which also
evidences gapless excitations.
The 1DTFIM predicts $1/T_1$$\sim$$T^{-0.75}$ at the QCP~\cite{Imai_PRX_2014},
while we observe a weakly $T$-dependent $1/T_1$ at $H_{C2}$.
The deviation may arise if our measurement missed the exact $H_{C2}$,
or if the temperature is not low enough.
It is worth noting that for fields just below $H_{C2}$, the rapid
drop of $1/T_1$ with decreasing temperature signals gapped excitations.
This agrees with the prediction in the 1DTFIM, and to our knowledge,
such a gap opening behavior has never been reported before.

We draw connections between our observations and the proposed
topological phase transitions
in BaCo$_2$V$_2$O$_8$~\cite{Faure_np__2018}.
The induced two-period staggered moments have been verified
by the neutron scattering study on BaCo$_2$V$_2$O$_8$~\cite{Faure_np__2018}.
We expect that they also appear in SrCo$_2$V$_2$O$_8$, given
their similar lattice and magnetic structures~\cite{Bera_PRB__2014}.
More evidence comes from the lower critical field along the [100] direction
in SrCo$_2$V$_2$O$_8$, which is consistent with a stronger induced
staggered field by the larger tilt angle of the O-Co-O bond
in this compound~\cite{Lorenz_PRB__2014}.
Although the materials are slightly different, the properties,
especially the gapped excitations, in both the ordered phase at
$H\ll H_{C}$ and the paramagnetic phase at $H\gg H_{C}$,
are the same.
However, when the 1D QCP is outside the 3D ordered phase, as in the case
of SrCo$_2$V$_2$O$_8$, gapless excitations emerge at the two QCPs,
since static linear confining potential is absent above $H_{C1}$.
Moreover, an exposed 1D QCP outside the 3D ordered phase at a relatively low field
promises an ideal approach for experimental investigation of the
genuine 1D quantum criticality. For future studies, by rotating the field
from [100] to [110] direction, two QCPs may be further separated.
This is indeed recently proved theoretically~\cite{ZhouHY_2019}.

In summary, our NMR study on SrCo$_2$V$_2$O$_8$
reveals two separated quantum critical points experimentally with a transverse field,
for the first time in this material.
Our results also demonstrate distinctive quantum critical behaviors
at the two QCPs, including the scaling behaviors of the
transition and crossover temperatures and the spin dynamics.
We propose that separate 1D and 3D QCPs are realized
at very low fields, attributed to novel effects of the effective staggered magnetic
field. Such a one-dimensional quantum critical point exposed
outside the three-dimensional magnetic ordered phase at a
low magnetic field opens a new avenue to access the
genuine quantum criticality of the one-dimensional
transverse-field Ising model experimentally.

The authors acknowledge discussions with Prof. Bruce Normand.
This work is supported by the Ministry of Science and Technology of China
(Grant Nos.~2016YFA0300504 and 2016YFA0300503), the National Natural
Science Foundation of China (Grant
Nos.~51872328, 21875249, and 11674392, and 11804221 ), the Fundamental Research
Funds for the Central Universities, the Research Funds of Renmin University
of China (Grant No.~18XNLG24), the Science and Technology
Commission of Shanghai Municipality Grant No.~16DZ226020, and
the Outstanding Innovative Talents Cultivation Funded Programs
2018 of Renmin University of China. JW acknowledges support from Shanghai city.

\end{document}